\documentclass[11pt,twoside]{article}
\usepackage{APN5proc}

\resetcounters

\begin{document}

\title{The Physical Characteristics of Binary Central Stars of Planetary Nebulae}
\author{Todd C. Hillwig
\affil{Valparaiso University, Valparaiso, IN USA 46383}}

\begin{abstract}
A number of efforts are underway to detect close binary stars in
planetary nebulae.  The primary goal of these studies is to determine
the binary fraction of central stars. The next stage is a detailed analysis
of the binaries to determine physical parameters for the systems.
These analyses can be combined with population synthesis models, common
envelope evolution models, and observed properties of nebulae to
further understand the impact of binarity on PN formation.  I
discuss the sample of known close binary central stars
in relation to other close binaries with a white dwarf, cataclysmic variables, supernova Ia progenitors, and double degenerate systems.\\ \noindent{\bf Keywords.}\hspace{10pt}Planetary Nebulae -- Binaries
\end{abstract}

\section{Introduction}

In our quest to understand the shaping of planetary nebulae (PNe) and how binary stars play a role in that process, it is important to characterize the discovered binary systems.  The masses of the component stars, binary separation, evolutionary state of both stars, stellar temperatures and radii, and observed inclination are all important pieces in understanding the connection between binary system and the ejected nebula.
In addition to understanding the physical parameters of the binary systems, it is important to know how the observed binary systems compare to what I will call post common envelope binaries (PCEBs)
that do {\it not} have a visible PN.  These field binaries can help us to determine if our sample of binary central stars of
PNe (CSPNe) represent the overall evolutionary sample, or if they comprise a
distinct subset.

Here I discuss the known binary star sample as a set, focusing on close binary CSPNe.  The sample of 36 close binary CSPNe comes from \citet*{dem08} and \citet{mis09a}.  I compare the current
sample to the known PCEB stars; specifically those with at least one white dwarf (WD) component.  I also look more closely at the class of binary stars observed in each case.  Below I discuss three distinct classes: 1) WD--main sequence (MS) detached systems, 2) WD--MS semi-detached systems or cataclysmic variables (CVs), and 3) double degenerates (DDs). 

\section{Comparison to Detached White Dwarf--Main Sequence PCEBs}

\begin{figure}
\begin{center}
\includegraphics[width=3.0in]{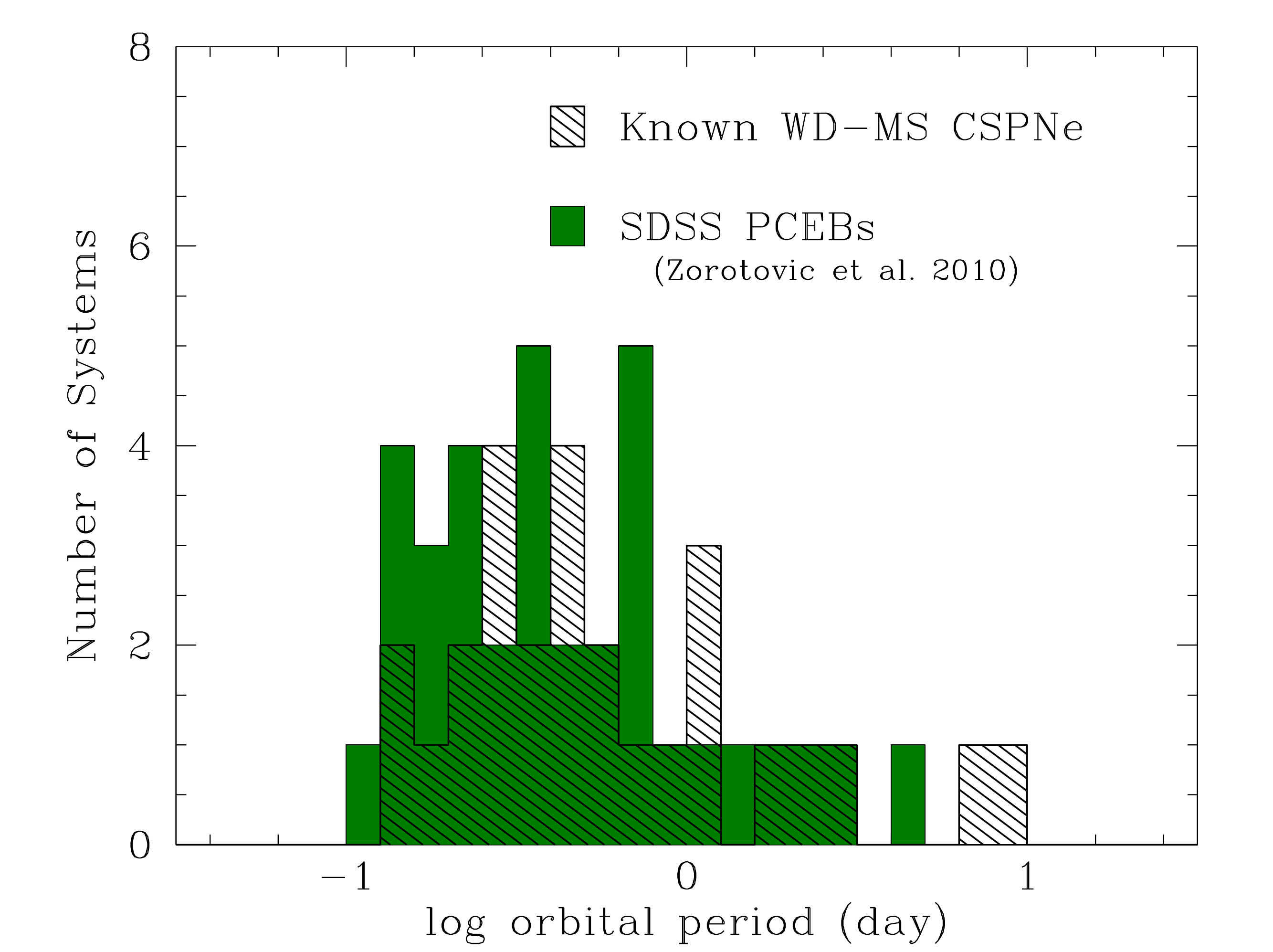}
\end{center}
\caption[Detached WD--MS histogram] {A histogram of the log of the orbital period in days of the known sample from SDSS
of detached WD--MS PCEBs ({\it solid green}) and the CSPNe ({\it hatched}).\label{pceb_histo}}
\end{figure}

While a significant amount of work has been done to discover and study MS--WD PCEB systems, there are still only a small number with known orbital periods.  The Sloan Digital Sky Survey (SDSS) has provided an excellent source for these PCEBs \citep*{zor10}.  Figure \ref{pceb_histo} shows a histogram of the WD--MS PCEB systems with known orbital periods from \citet{zor10} plotted as a function of the log of the orbital period in days along with the distribution of WD--MS CSPNe with known orbital period.  
The distributions of field and PN close WD--MS binaries appear to be in good agreement.  The two CSPNe with orbital periods close to ten days suggest a possible long orbital period tail,  but the total numbers will have to increase significantly before such a difference between the two could be confirmed.  If there is a tail, it makes up only a small fraction of the total systems.

Also apparent is that the majority of systems in both distributions have orbital periods $< 1$ day.  In the case of the CSPNe, which have been discovered through photometric variability, it is possible that observational selection effects could play a role and that longer period systems do exist, though \citet{dem08} demonstrate that this is unlikely.  Such a selection effect is unlikely for the PCEB distribution however, as these systems were discovered via radial velocity variability and the SDSS data are of high enough resolution to allow detection of significantly longer orbital periods.

\section{Comparison to Cataclysmic Variables}

Recently, the classical nova GK Per \citep{sco94} and recent nova V458 Vul \citep{rod10} have had reports of faint PNe surrounding them.  Also, observations of the binary CS of Hen 2-428 suggest that it may be a semi-detached system \citep{san10}.

\begin{figure}
\begin{center}
\includegraphics[width=3.0in]{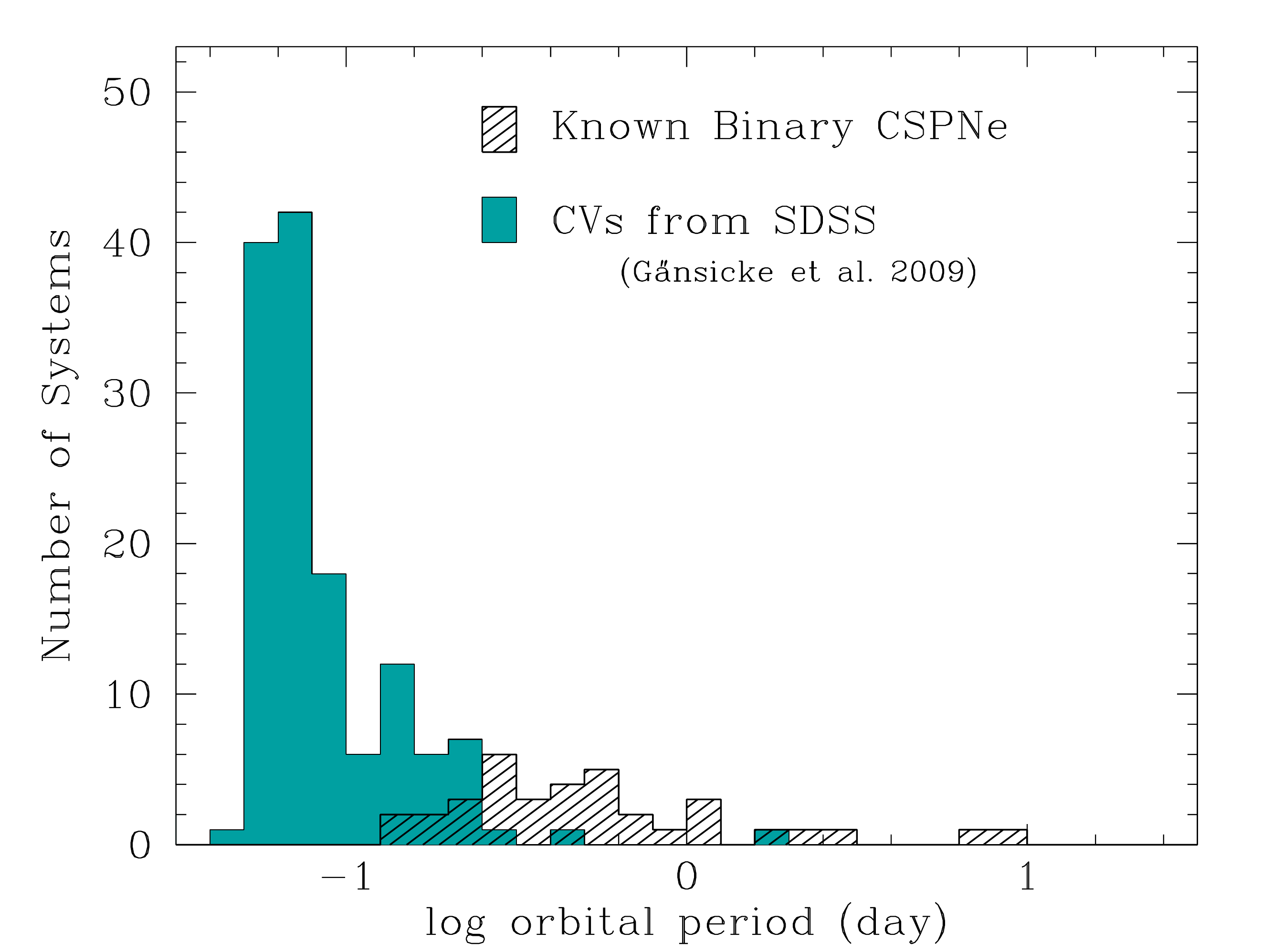}
\end{center}
\caption[CV histogram] {A histogram of the log orbital period in days of the known sample of CVs from SDSS
({\it solid blue}) and all close binary CSPNe ({\it hatched}).\label{cv_histo}}
\end{figure}

In Figure \ref{cv_histo} we have plotted the complete distribution of close binary CSPNe in a histogram along with the SDSS sample of 137 CVs from \citet{gan09}, which is the best available unbiased sample of CVs (though perhaps with low sensitivity to long
period CVs).
The two samples show a very small overlap.  Furthermore, the period region in which they do overlap is above the CV period gap where a semi-detached state would only occur for relatively high mass secondary stars (typically $> 0.5$ M$_\odot$).  Based on these distributions we expect that very few CSPNe systems would have evolved directly into contact, and that most systems reach a semi-detached state after magnetic or gravitational braking has reduced their orbital period.  This result is in agreement with the small number of known CSPNe in semi-detached systems and with the conclusions of \citet{gan09} for
PCEBs in general.

\section{Double Degenerate Binary Central Stars}

In the context of CSPNe, I will use ``double degenerate'' to mean a binary system in which the two stars are either WDs or pre-WDs (technically not degenerate yet, but will become so).  One reason to understand the fraction and classification of DD binary CSPNe is their possible relationship to type Ia supernovae (SNIa).  Another is that the fraction of helium WDs versus carbon-oxygen WDs combined with population synthesis calculations can help us determine typical evolutionary channels for CSPNe.

To date three DD CSPNe have been firmly identified in the literature: the CSs of Abell~41 \citep{shi08}, PN~G135.9+55.9 \citep{tov10}, and NGC~6026 \citep{hil10}.  To this list we can add the CSs of Hen~2-428 and V458 Vul \citep{san10,rod10}.   The calculated masses for each component in all of the systems are too massive to be He WDs (for which M $\lesssim 0.46$ M$_\odot$).  A possible exception is the CS of NGC~6026 where the published mass range does include He WD mass values.  The dominance of CO WDs contradicts models in which most WDs in DDs will be He WDs \citep[e.g.][]{nel01}.  Therefore, either the four studied systems represent a significant departure from the true sample or CSPNe are intrinsically different than field WDs.  If the small sample does not represent the true distribution of WDs in PNe it may either be due to a random selection of only high mass systems or due to an observational bias toward the higher mass CO WDs.  

\begin{figure}
\begin{center}
\includegraphics[width=3.0in]{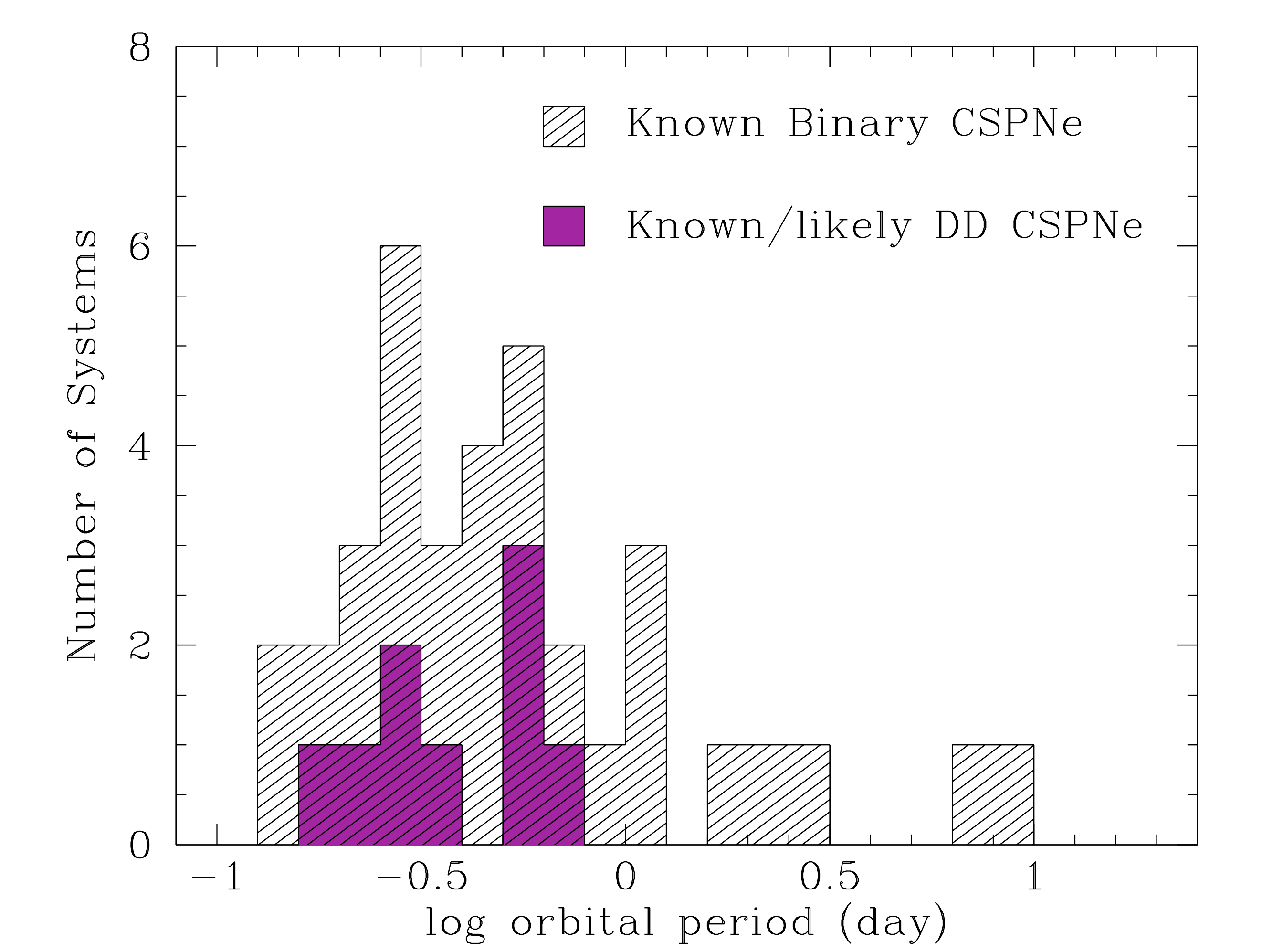}
\end{center}
\caption[DD histogram] {A histogram of in log orbital period of the known sample of CSPNe
({\it hatched}) with known or suspected DD systems ({\it solid purple}).\label{dd_histo}}
\end{figure}

Each of the four known DD CSPNe has a light curve dominated by
ellipsoidal variability.  These are the only well-studied binary CSPNe showing this type of light curve.  Through binary system modeling we find that it is very difficult to have a light curve dominated by ellipsoidal variability in non-DD systems.  Also, light curves with two eclipses of nearly equal depth (nearly equal temperature stars) will also be DD systems.  In the sample of known close binary CSPNe \citep{dem08,mis09a} we find that nine of the 36 systems have light curves dominated by ellipsoidal variability or show two nearly equal-depth eclipses (see histogram in figure \ref{dd_histo}).
So one-quarter of the known binary CSPNe are known or suspected double-degenerates.  From a statistical standpoint this suggests again a possible difference between DDs in CSPNe and those in field stars.  The Supernova Ia Progenitor Survey \citep[SPY][]{nap03} finds from a radial velocity survey that 15\% of {\it all} field WDs are DDs, whereas for an assumed close binary fraction of 20\% we find only 5\% of {\it all} CSPNe to be DDs.

The distribution and classification of DD CSPNe suggests that PNe may harbor a different distribution, with more massive components, than field PCEBs.  If after enlarging the sample this still holds, CSPNe would provide a good test ground for SN Ia searches and for exploring DD evolutionary channels in general.

\acknowledgements This research was supported in part by NASA through the American Astronomical Society's Small Research Grant Program.


\begin{thebibliography}

\bibitem[De Marco et al.(2008)De Marco, Hillwig, \& Smith]{dem08}
     De Marco, O., Hillwig, T., \& Smith, A.J. 2008, \aj, 136, 323

\bibitem[G\"{a}nsicke et al.(2009)]{gan09}
     G\"{a}nsicke, B. et al. 2009, \mnras, 397, 2170

\bibitem[Hillwig et al.(2010)]{hil10}
     Hillwig, T. C., Bond, H. E., Af\c{s}ar, M., \& De Marco, O. 2010, \aj, 140, 319

\bibitem[Miszalski et al.(2009)]{mis09a}
     Miszalski, B., Acker, A., Moffat, A.F.J.,  Parker, Q.A., \& Udalski, A. 2009,
     \aap, 496, 813

\bibitem[Napiwotzki et al.(2003)]{nap03}
     Napiwotzki, R. et al. 2003, {\it Messenger}, 112, 25

\bibitem[Nelemans et al.(2001)]{nel01}
     Nelemans, G., Yungelson, L. R. , Portegies Zwart, S. F., \& Verbunt, F. 2001 \aap, 365, 491

\bibitem[Rodr{\'{\i}}guez-Gil et al.(2010)]{rod10} 
Rodr{\'{\i}}guez-Gil, P., et al.\ 2010, \mnras, 407, L21

\bibitem[Santander-Garc{\'{\i}}a et al.(2010)]{san10}
     Santander-Garc{\'{\i}}a, M. et. al. 2010, {\it Asymmetric Planetary Nebulae 5}, published by Ebrary, Palo Alto,
     California, USA.  A. A. Zijlstra, I. McDonald, and E. Lagadec (eds.), this proc.

\bibitem[Scott et al.(1994)]{sco94} 
     Scott, A.~D., Rawlings, J.~M.~C., \& Evans, A.\ 1994, \mnras, 269, 707

\bibitem[Shimanskii et al.(2008)]{shi08}
    Shimanskii, V. V., Borisov, N. V., Sakhibullin, N. A., \& Sheveleva, D. V. 2008, ARep, 52, 479

\bibitem[Tovmassian et al.(2010)]{tov10}
    Tovmassian, G., et al. 2010, \apj, 714, 178

\bibitem[Zorotovic et al.(2010)]{zor10}
     Zorotovic, M., Schreiber, M. R., G\"{a}nsicke, B. T., \& Nebot G{\'{o}}mez-Mor{\'{a}}n, A. 2010, arXiv:1006.1621 

\end{thebibliography}

\end{document}